\documentstyle[11pt,psfig]{article}
\begin{document}
\LARGE
\begin{center}
{\bf The influence of dynamical friction on the collapse 
of spherical density pertubation. }
\large
~\\
~\\
{\bf A. ~Del Popolo, ~M. ~Gambera, ~V. ~Antonuccio-Delogu }\\
~\\
Istituto di Astronomia dell'Universit\`a di Catania, \\
\&\\
Osservatorio Astrofisico di Catania\\
Citt\`a Universitaria, Viale A.Doria, 6 - Italy\\

\large

\begin{abstract}
We solve numerically the equations of motion for the collapse 
of a shell of baryonic matter falling into 
the central regions of a cluster of galaxies. 
taking into account of the presence of the   
substructure inducing dynamical friction. The 
evolution of 
the expansion parameter a(t) of the perturbation is calculated in 
spherical systems. The effect of dynamical friction 
is to reduce the binding radius and the toatal mass accreted by 
the central regions. 
Using a peak 
density profile given by Bardeen et al. (1986) we show how the 
binding radius 
of the perturbation is 
modified by dinamical friction. We show how dynamical 
friction modifies the collapse parameter of the perturbation slowing down 
the collapse.
\end{abstract}
~\\
~\\
\end{center}

\large
\begin{center}
{\bf To be published:}\\

{\sc Memorie della Societ\`a Astronomica Italiana}\\
\end{center}

\newpage

\large
\begin{flushleft}
 
{\bf 1. Introduction}
\end{flushleft} 


%

\noindent
In the most promising cosmological scenarios structure formation 
in universe is generated through the growth and collapse 
of primeval density 
perturbations originated from quantum fluctuations (Guth \& Pi 1982; 
Hawking 1982; Starobinsky 1982; Bardeen et al. 1986) in an inflationary 
phase of early universe. 
perturbations is due to gravitational instability. 
The statistics of 
density fluctuations originated in the inflationary era are Gaussian and 
it can be expressed entirely 
in terms of the power spectrum of density fluctuations: 
\begin{equation}
P( k) = < |\delta_{{\bf k}}|^{2}> 
\end{equation}
where 
\begin{equation}
\delta_{{\bf k}} =\int d^{3} k exp(-i {\bf k x}) \delta({\bf x})
\end{equation}
\begin{equation}
\delta({\bf x}) = \frac{ \rho ({\bf x}) - \rho_{b}}{ \rho_{b} }
\end{equation}
and $ \rho_{b} $ is the mean background density. 
In biased structure formation theory it is assumed that cosmic structures 
of linear scale $ R$ form around the peak of density field, 
$  \delta( {\bf x})$, smoothed on the same scale. 
Density perturbations evolve towards non-linear regime bacause of 
gravitational instability and it breaks away from general expansion 
at: 
\begin{equation}
t_{m} = \left[\frac{ 3 \pi}{32 G \rho_{b}} ( 1 +\overline{\delta})
\right]^{1/2} (1+z)^{3/2}
\end{equation}
where $ z$ is the redshift, $ \overline{\delta}$ is the overdensity 
within $ r$. When $ \overline{\delta} \simeq 1$ the density 
perturbation begins to recollapse. The collapse time, $T_{c0}$, 
depends on the characteristic of initial fluctuation field as 
average overdensity and on the environment in which the perturbation 
is embedded. This last features depends on the cosmological scenario . 
The mean characteristics of the structure of the structures which 
form around the peak of the density field depend on 
the spectrum, $ P(k)$, which in turn depends on the matter that dominates 
the universe (CDM, WDM, HDM). In this context the most succesful 
model is the biased CDM model (Liddle \& Lith 1994) based on a scale 
invariant spectrum of density fluctuations growing under 
gravitational instability and on the hypothesis that the universe 
is dominated by cold dark matter.\\ 
A very simple model for 
accretion of matter in cluster of galaxies was first investigated 
by Gunn \& Gott (1972). In their accretion model the density 
profile is given by:
\begin{eqnarray}
\rho_{i} = \rho_{ei} +( \rho_{ci} +\rho_{+} -\rho_{ei}) \frac{ R_{i}^{3}}
{ r_{i}^{3}} \hspace{1.0cm} r> R_{i} \nonumber \\
\rho_{ci}+\rho_{+}  \hspace{1.0cm} r_{i} \leq R_{i} 
\end{eqnarray}
where $ \rho_{ci}$ and $ \rho_{ei}$ are the critical and 
external density, $ \rho_{+}$ gives the density surplus present 
in the radius $ R_{i}$ respect to the critical density. The perturbation 
with $ \rho > \rho_{ci} $ reach a 
maximum radius $ r_{m} =\frac{x}{\overline{\delta}}$ 
at a time: 
\begin{equation}
T_{c0}/2 =\frac{ \pi}{ H_{i}} 
\frac{(1 +\overline{\delta})}
{\overline{\delta^{3/2}}}
\label{eq:beef}
\end{equation}
where $ \delta$ is the overdensity in the radius $ r $ and $ H_{i} $ 
is the Hubble parameter at time $ t_{i} $. After $ T_{c0}/2 $ 
the matter collapse and its infall is radial. The perturbation 
needs a time $T_{c0} $ for the total collapse. Gunn \& Gott 
model 
is a oversimplification of the perturbation evolution.   
In Gunn \& Gott 
model there are no tidal interactions among the shell of baryonic matter
and the external density perturbations and it is supposed that there is 
no substructure, (collapsed objects of lenght less than that of 
main perturbation). One of the features of CDM models is an abundant 
production 
of substructure. In this scenario structure formation goes from 
bottom to up through gravitational clustering, merging and violent 
relaxation of small scale substructure (White \& Rees 1978). So the matter 
inside a given region is clumped in a hierarchy of objects of various 
dimensions. The time a given lenght goes nonlinear can be obtained from 
the condition on mass spectrum, $ \sigma_{0} (M)$: 
\begin{equation}
\sigma_{0} (M) = 1= 1 +z_{nl} 
\end{equation}
where: 
\begin{equation}
\sigma_{0} (M) = \frac{ 1}{ 2 \pi^{2}} \int P(k) k^{2} W( k R_{f}) d k
\end{equation}
beeing $ R_{f} $ the filtering scale and $ W( k R_{f})$ the window function: 
\begin{equation}
W( k R_{f}) = \frac{ 3[ sin( k R_{f}) - k R_{f} cos( k R_{f})]}{k R_{f}^{3}}
\end{equation}
So a subgalactic scale collapse at a redshift: 
\begin{equation}
1+z_{nl} =\frac{30}{b}
\end{equation}
(Silk \& Stebbins 1993), where $ b $ is the biasing parameter. 
In particular a perturbation of $ 10^{15} M_{\odot} $ 
collapses at a redshift $ z \simeq   0.02$ while perturbations 
in the range $ 10^{6} M_{\odot} - 10^{9} M_{\odot}$ collapse almost at 
the same $ z \simeq 18 $ because the mass variance in this region 
varies only of a factor 3 (Rees 1986). A part of this last 
perurbations has a cross section too little for gravitational 
merging to be destroied (Rees 1986). Therefore a mass perturbation 
of $ 10^{6} M_{\odot} - 10^{9} M_{\odot}$ can survive untill the 
cluster enters nonlinear phase at $ z\simeq 0.02$. Because of 
the presence of substructure in a cluster Gunn \& Gott model needs 
a revision. In fact substructure acts as a source of 
stochastic fluctuations in 
gravitational field and induces 
dynamical friction (Antonuccio \& Colafrancesco 1994) and this 
produces a modification of the motion of shells of baryonic matter 
in a density perturbation. In particular dynamical friction 
delays the collapse of low density perturbations 
($\overline{\delta} \simeq 0.01$) (Antonuccio \& 
Colafrancesco 1994). 
The plan of the paper is as follows. In section 2
by numerical integration of equations of motion of a shell of 
baryonic matter made of galaxies and substructrure of  
$ 10^{6} M_{\odot} - 10^{9} M_{\odot}$ we show  how 
the expansion parameter of the same varies with time. In section 3 we show
how dynamical friction affects the binding radius of a cluster. 
~\\

\begin{flushleft}
{\bf 2. Modification of the expansion parameter of a shell.}
\end{flushleft}

\noindent
The equation of motion of a shell of baryonic matter around a 
maximum of the density field, neglecting tidal interactions 
and substructure, can be expressed in the form:
\begin{equation}
\frac{d^{2} r}{d t^{2} }= -\frac{G M}{r^{2}(t)} \label{eq:pee}
\end{equation}
(Peebles 1980, eq. 19.9), where $ M$ is the mass enclosed in the proper 
radius $ r(t)$. Using Gunn \& Gott notation the proper radius 
can be written as: 
\begin{equation}
r(r_{i},t) = 
a( r_{i}, t) r_{i} 
\end{equation}
where $ r_{i} $ is the initial radius, $ a(r_{i},t)$ is the expansion  
parameter of the shell. At the initial time $ t_{i}$ the initial 
condition is given by $ a(r_{i},t_{i}) =1$. \\
In presence of substructure it is necessary 
to modify the equation of motion Eq. (\ref{eq:pee}) because of the graininess 
of mass distribution in the system, due to substructure. In a material 
system gravitational field can be decomposed into an average field, 
$ {\bf F}_{0}(r)$, generated from the smoothed out distribution 
of mass, and a stochastic component, $ {\bf F}_{stoch}(r)$, generated 
from the fluctuations in number of the neighbouring particles. 
The stochastic component of the gravitational field is 
specified assigning a probability density, $ W( {\bf F})$, (Chandrasekhar \& 
von Neumann 1942). In a infinite Homogeneous unclustered system 
$ W( {\bf F})$ is given by Holtsmark distribution 
(Chandrasekhar \& 
von Neumann 1943) while in inhomogeneous and clustered systems $ W({\bf F})$ 
is given by Kandrup (1980) and Antonuccio \& Barandela (1992) respectively. 
The stochastic force, $ {\bf F}_{stoch}$, in a self-gravitating 
system modifyes the motion of particles as it is done by a frictional force. 
In fact a particle moving faster than its neighbours produces a deflection 
of their orbits in such a way that average density is greater in 
the direction opposite to that of motion causing a slowing down 
in its motion. The modified equation of motion of a shell of baryonic 
matter can be written in the form:
\begin{equation}
\frac{d^{2} {\bf r}}{d t^{2} }= -\frac{G M}{r^{2}(t)} -\eta {\bf v} 
\label{eq:pees}
\end{equation}
(Langevin 1902, Saslaw 1985) where $ \eta$ is the coefficient of 
dynamical friction. Supposing that there are no correlations 
among random force and their derivatives we have: 
\begin{equation}
\eta = \frac{\int d^{3} F W(F) F^{2} T(F) }{ 2 < v^{2} >}
\end{equation}    
(Kandrup 1980), where $ T(F)$ is the average duration of a random 
force impulse, $ W(F)$, is the probability distribution 
of stochastic force that for a clustered system is given 
by Antonuccio \& Barandela (1992). Equation 
(\ref{eq:pees}) in terms of $ a( r_{i} ,t)$ and 
\begin{equation}
\rho(r_{i}, t) = \frac{3 M}{4 \pi a^{3}(r_{i},t) r_{i}^{3} }= 
\frac{\rho (r_{i}, t_{i})}{ a^{3} (r_{i}, t_{i})} = \rho_{ci}( 1+ 
\overline{\delta_{i}})
\end{equation}
can be written as:
\begin{equation}
\frac{d^{2} a}{d t^{2} }= -
\frac{4 \pi G \rho_{ci}( 1+\overline{\delta_{i}})}{a^{2}(t)} -
\eta \frac{ d a }{ d t} 
\label{eq:peesl}
\end{equation}
where $ \rho_{ci} $ is the background density a time $ t_{i} $ 
and $ \overline{ \delta_{i}}$ is the overdensity within $ r_{i}$. 
Using the parameter $ \tau= \frac{t}{ T_{c0} }$ the Eq. (\ref{eq:peesl})
may be written in the form:
\begin{equation}
\frac{d^{2} a}{d \tau^{2} }= -
\frac{4 \pi G \rho_{ci}( 1+\overline{\delta_{i}})}{a^{2}(\tau)} T_{co}^{2}-
\eta T_{c0} \frac{ d a }{ d \tau} 
\label{eq:pessr}
\end{equation}
Referring to the calculation of $ \eta $, given by Antonuccio 
\& Colafrancesco (1994) and using equation (\ref{eq:pessr})
the time evolution of the expansion parameter, $ a(\tau)$, of the shell can be 
obtained. 
Equation (\ref{eq:pessr}) can be solved looking for an asymptotic expansion. 
This was made by Antonuccio \& Colafranceasco (1994). In their 
paper they gave only an expression for the collapse time $ T_{c}$. Equation 
(\ref{eq:pessr}) can be also solved numerically. 
To this aim we used a Runge-Kutta integrator 
of $ 4^{th}$ order. We studied the motion of a shell of low density, 
$ \overline{\delta} =0.01$, typical of a perturbatoon present 
in the outskirts of a cluster of galaxies. The initial conditions 
were chosen remembering that in expansion phase the shell moves with 
Hubble flow and that at the maximum of expansion the initial 
velocity is zero. 
In figure 1, we show the expansion parameter $a(\tau) $ versus 
 $\tau$ both when the dynamical friction is taken into account and
when it is absent.
\begin{figure}[ht]
\psfig{file=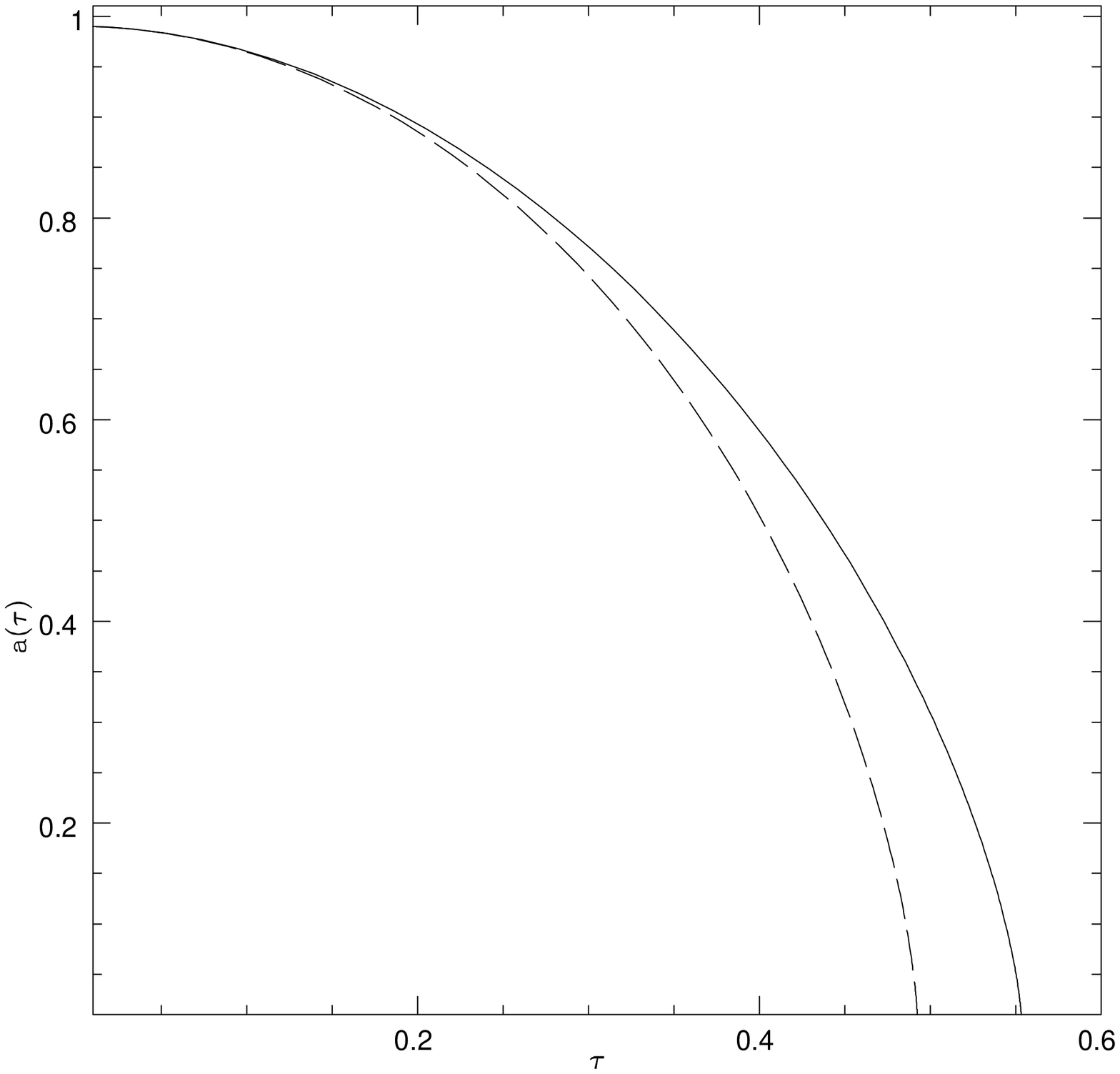,width=12cm}
\caption[h]{Temporal evolution of the expansion parameter $ a(\tau)$  of a shell of 
matter made of galaxies and substructure versus $\tau$. The dashed line  shows
$a(\tau)$  when 
dynamical friction is absent, while the solid line  is the same 
when dynamical friction is taken into account.
We assume a cluster radius of $R_{cl} = 5 h^{-1} Mpc $, a central overdensity 
$\overline{ \delta} =0.01$ and 
a total number of peaks of substructure $ N_{tot} = 10^{3}$. }
\end{figure}
As we can see dynamical friction 
slows down the collapse of the shell of matter in agreement with the 
analytic calculation of the collapse time, $ T_{c} $, of a shell 
in which dynamical friction effect is taken into account 
given by Antonuccio \& Colafrancesco (1994). 

\begin{flushleft}
{\bf 3. Binding radius of a cluster in presence of dynamical 
friction.} 
\end{flushleft}

In biased galaxy formation theory structures form around the 
local maxima of the density field. Every density peak binds 
a mass $ m$ that can be calculated when we know the binding radius 
of the density peak. The radius of the bound region 
for a chosen density profile $\overline{\delta}( r)$ may be 
obtained in several ways. A first criterion is statistic. The binding 
radius of the region, $ r_{b}$, is given by the solution of the equation:  
\begin{equation}
< \overline{\delta} (r)> = 
< ( \overline{\delta} -<\overline{\delta}>)^{2}>^{1/2}
\end{equation}
(Ryden 1988). At radius $ r << r_{b}$ the motion of particles is 
predeminant toward the peak while when $r >> r_{b} $ the particle 
is not bound to the peak. Another criterion 
that can be used is dynamical. It supposes that the binding radius 
is given by the condition that a shell collapse in a time, $ T_{c}$, 
smaller than the age of the universe $ t_{0}$: 
\begin{equation}
T_{c} (r) \leq t_{0} \label{eq:temp}
\end{equation} 

\begin{figure}[ht]
\psfig{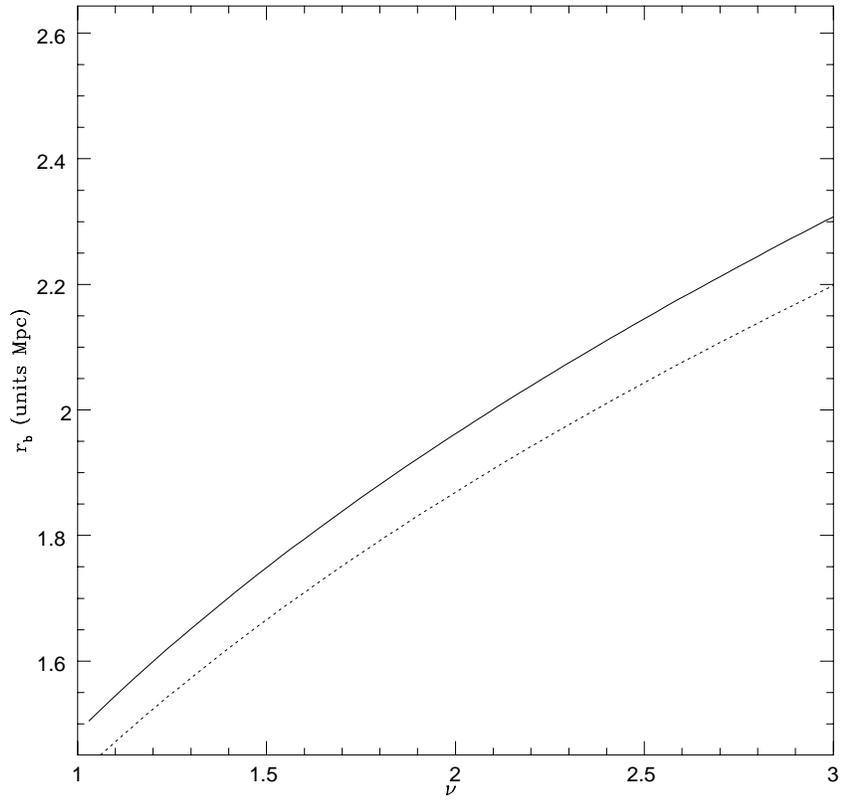}
\caption[h]{Variation of the binding radius $ r_{b} $ with $ \nu $.
The solid line is the binding radius in absence of dynamical friction,
while the dashed line is the same as in presence of dynamical friction.
The filtering radius used is , $ R_{f} = 1 \hspace{0.1cm} Mpc $ }
\end{figure}

(Hoffmann \& Shaham 1985). This last criterion, differently from 
the previous one, contains some prescriptions particularly connected with the 
physics of the collapse process of a shell. For this reason 
we used it to calculate the binding radius. The time of collapse, $ T_{c}(r)$, 
at radius $ r$ can be obtained solving numerically equation 
(\ref{eq:peesl})
for different values of $ \overline{\delta_{i}}$ from  
a given density profile $ \overline{\delta}(r)$. We use the average density 
profile given by Bardeen et al. (1986): 
\begin{equation}
\delta(r)= A \left\{ \frac{\nu \xi(r)}{ \xi(0)^{1/2}}- 
\frac{\theta( \nu \gamma, \gamma)}{\gamma \xi(0)^{1/2}(1-\gamma^{2})}
\left[\gamma^{2} \xi(r) +\frac{ R_{\ast}^{2} 
\bigtriangledown^{2} \xi(r)}{3}\right]\right\} \label{eq:dens}
\end{equation} 
where A is a constant given by the normalization of the 
perturbation spectrum, $ P( k)$,  $ \nu = \frac{\delta(0)}{\sigma_{0}(M)} $, 
$ \xi(r)$ is the correlation function of two points, $ \gamma$ and $ R_{\ast}$ 
two constants obtainable from the spectrum (see Bardeen et al. 1986) and 
finally $ \theta ( \gamma \nu, \gamma)$ is a function given in 
the quoted paper (eq. 6.14). Given the average density 
profile the average density inside the radius $ r$ in  a spherical 
perturbation is given by: 
\begin{equation}
\overline{\delta} = \frac{3}{r^{3}} \int_{0}^{r} d x \delta(x) x^{2} 
\end{equation}
We calculated the time of collapse, $ T_{c0}(r)$, using Eq.~(\ref{eq:beef}) 
with the density profile given in Eq.~(\ref{eq:dens}).
We repeated the calculation 
of $ T_{c0} (r)$ for 
$ 1.5 <\nu< 3 $ and we applied the condition given in Eq. (~\ref{eq:temp}) 
to the curves $ T_{c0} (r)$ previously obtained. The result is the plot 
in Fig. ~2 for the binding radius $ r_{b} $ versus $ \nu$. 

\begin{flushleft}
{\bf 4. Conclusion}
\end{flushleft}

In the first part of this paper we show how the dynamical friction
slow down the collapse of a spherical density perturbation.
The effect grows with increasing of the coefficient of dynamical
friction $\eta$. In the second part, we show how dynamical friction
reduces the binding radius (see fig.2).






\end{document}